\documentclass[prl,numbers,amsmath,amssymb,aps]{revtex4}

\usepackage{graphicx}
\usepackage{amssymb}
\usepackage{dcolumn}
\usepackage{bm}
\newcommand{\gbf}[1]{\mbox{\boldmath$#1$}}

\begin{document}

\title{White-light parametric instabilities in plasmas}
\author{J. E. Santos\footnote{Now at: Department of Applied Mathematics and Theoretical Physics, Centre for Mathematical Sciences, University of Cambridge, Wilberforce Road, Cambridge CB3 0WA, England}}
\author{L. O. Silva}
\email{luis.silva@ist.utl.pt}
\affiliation{GoLP/Centro de F\'isica dos Plasmas, Instituto Superior T\' ecnico, 1049-001 Lisboa, Portugal}
\author{R. Bingham }
\affiliation{Rutherford Appleton Laboratory, Chilton, Didcot, Oxon OX11 0QX, United Kingdom}

\begin{abstract}
Parametric instabilities driven by partially coherent radiation in plasmas are described by a generalized statistical Wigner-Moyal set of equations, formally equivalent to the full wave equation, coupled to the plasma fluid equations. A generalized dispersion relation for Stimulated Raman Scattering driven by a partially coherent pump field is derived, revealing a growth rate dependence, with the coherence width $\sigma$ of the radiation field, scaling with $1/\sigma$ for backscattering (three-wave process), and with $1/\sigma^{1/2}$ for direct forward scattering (four-wave process). Our results demonstrate the possibility to control the growth rates of these instabilities by properly using broadband pump radiation fields.
\end{abstract}

\date{\today}

\maketitle


Parametric instabilities are pervasive in many fields of science, associated with the onset of nonlinear and collective effects such as solitons, vortices, self-organization, and spontaneous ordering. Recent developments in light sources and laser technology continue to reveal novel features of the parametric instabilities, for instance in nonlinear optics, with the recent experimental discovery of white light solitons \cite{bib:mitchel}, or in plasma physics, in the realm of relativistic nonlinear optics \cite{bib:mori}.
The standard theoretical approach to study parametric instabilities is based on a coherent wave description which is clearly limited because, in most systems, waves are only partially coherent, with incoherence either inherently induced by fluctuations, or induced by external passive systems (e.g. random phase plates in inertial confinement fusion (ICF)). 
Recent theoretical work in nonlinear optics, triggered by the work of Segev and co-workers \cite{bib:mitchel}, led to the development of  techniques capable of describing the propagation and the modulation instability of partially coherent/incoherent "white" light in nonlinear media 
\cite{bib:buljan}. The critical underlying assumption of all these models is the paraxial wave approximation, valid in transparent media for radiation beams not tightly focused, which reduces the problem of electromagnetic wave propagation in dispersive (and diffractive) nonlinear media to the search of a forward propagating solution, formally described by the nonlinear Schr\"odinger equation.    
While in nonlinear optics, and for the conditions studied so far, such approximation is clearly valid, in plasma physics it is not. The instabilities associated with the partially reflected backscattered radiation \cite{bib:kruer} are critical in many laser-plasma and astrophysical scenarios \cite{bib:thompson}, and the paraxial approximation has limited applicability even in the description of forward scattering instabilities driven by ultra intense lasers in underdense/transparent plasmas \cite{bib:mora}.    

Inclusion of bandwidth/incoherence effects in laser driven parametric instabilities in plasmas and in three-wave processes is a long-standing problem \cite{bib:zasvaleo, bib:tsytovich}, because incoherent pumps can decrease the growth of the laser driven instabilities in ICF, Fast Ignition, or novel laser amplification schemes \cite{bib:fisch}. 
The difficulty resides in the lack of an appropriate theoretical framework where a statistical description of the radiation is natural. 
The Wigner-Moyal (WM) formalism of quantum mechanics \cite{bib:liboff} provides a natural path to build such a statistical description of the radiation \cite{bib:tappert}. However, it is important to point out that the standard WM approach is valid only for Schr\"odinger-like systems, where backscattered radiation is neglected. Previous attempts \cite{bib:pk} have only been able to describe direct forward stimulated scattering, failing to describe in a general way parametric processes driven by "white" light in plasmas. To overcome this difficulty, we have recently developed a generalized WM statistical theory of radiation \cite{bib:santos}, or generalized photon kinetics (GPK), formally equivalent to the full wave equation, valid for partially coherent electromagnetic wave propagation in nonlinear dispersive and diffractive media. 

In this Letter, GPK is employed to derive the general dispersion relation for Stimulated Raman Scattering (SRS) driven by a spatially stationary radiation field with arbitrary statistics, thus valid for all ranges of coherence of the pump field {\it i.e.} from a coherent plane wave pump to an incoherent pump. In this dispersion relation both three-wave processes and four-wave processes are considered, and for a plane wave pump field the standard results are recovered \cite{bib:kruer,bib:mora,bib:mckinstrie}. Analytical results are derived for different regimes of SRS and wavenumber ranges, showing universal decays of the growth rate with the bandwidth $\sigma$, with a $1/\sigma$ dependence for backscattering, and a slower decay, with $1/\sigma^{1/2}$, for forward scattering. Numerical solutions confirm the theoretical predictions, highlighting some of the most important consequences of white light in SRS.   

In our discussion, we use normalized units, such that length is normalized to $c/\omega_{p0}$, time to $1/\omega_{p0}$, mass and charge to the electron mass $m_e$ and the electron charge $e$, respectively, and where $c$ is the velocity of light in vacuum, and $\omega_{p0}= \left( 4 \pi e^2 n_{e0} /m_e c^2 \right)^{1/2}$. We model the plasma as a cold uniform electron fluid with a fixed ion density $n_{i0}=n_{e0}$ (in normalized units $n_{i0}=n_{e0}=1$). The normalized vector potential of the circularly polarized pump field $\mathbf{a}_p=e\mathbf{A}_p/m_e c^2$ is described by $\mathbf{a}_p(\mathbf{r},t)=2^{-1/2}(\hat{z}+i\hat{y}) a_0 \int d \mathbf{k} A(\mathbf{k})  
\exp{[i(\mathbf{k} \cdot \mathbf{r}-\omega(\mathbf{k}) t)]}$, 
where $\omega(\mathbf{k})$ is the dispersion relation for plane circularly polarized monochromatic waves in a uniform plasma, $\omega(\mathbf{k}) = \left( \mathbf{k}^2  + 1/\gamma_0\right)^{1/2}$, and $A(\mathbf{k})$ can include a stochastic phase dependence $\psi(\mathbf{r},t)$, as $A(\mathbf{k})=\hat{A}(\mathbf{k}) \exp \left(i \psi(\mathbf{r},t)\right)$.
The only restriction on the form of $\mathbf{a}_p(\mathbf{r},t)$ is that the Klimontovich statistical average of the two point correlation function, $\langle \mathbf{a}_p^*(\mathbf{r}+\mathbf{y}/2,t) \cdot \mathbf{a}_p(\mathbf{r}-\mathbf{y}/2,t) \rangle = a_0^2 m(\mathbf{y})$, is independent of $\mathbf{r}$ with $m(0)=1$, and $|m(\mathbf{y})|$ is bounded between 0 and 1 {\it i.e.} the field is spatially stationary. This restriction is introduced only because of the perturbation technique we are employing; the formalism described here is valid for any field dependence. When the uniform plasma is irradiated by the electromagnetic field, 
described by the normalized vector potential $\mathbf{a}=\mathbf{a}_p +\tilde{\mathbf{a}}$, the normalized electron density $n_e= 1 + \tilde{n}$ satisfies \cite{bib:decker}:
\begin{equation}
\left(\partial_t^2+\frac{1}{\gamma_0}\right)\tilde{n}=\frac{1}{\gamma_0^2}\nabla_\mathbf{r}^2(\langle\mathrm{Re}\left[\mathbf{a}_p\cdot\tilde{\mathbf{a}}\right] \rangle)
\label{eqn:dens}
\end{equation}
with $\gamma_0 = \sqrt{1+\langle\mathbf{a}_p \cdot \mathbf{a}^*_p\rangle} =\sqrt{1+a_0^2}$, 
where $\langle \cdots \rangle$ denotes a statistical average, and $\tilde{\, }$ denotes first-order quantities. 
Usually, the driving term on the right hand side of Eq.~(\ref{eqn:dens}) is described with the standard approach based on the wave equation for the vector potential \cite{bib:kruer,bib:decker}. However, this technique does not allow for the study of "white" light parametric instabilities. GPK  can address the general two mode problem. The radiation field $\mathbf{a}$ is described by two fields  $\gbf{\phi},\gbf{\chi}=(\mathbf{a}\pm i \sqrt{\gamma_0}\partial_t \mathbf{a})/2$, in terms of which the wave equation can be written as two coupled Schr\"odinger equations \cite{bib:santos}. Introducing four real phase-space densities  $W_0=W_{\phi\phi}-W_{\chi\chi}$, $W_1=2 \mathrm{Im}\left[W_{\phi\chi}\right]$, $W_2=2 \mathrm{Re}\left[W_{\phi\chi}\right]$, $W_3=W_{\phi\phi}+W_{\chi\chi}$, where the Wigner transform $W_{\mathbf{f}\cdot\mathbf{g}}$ is defined as $W_{\mathbf{f}\cdot\mathbf{g}}(\mathbf{k},\mathbf{r},t)=\left(\frac{1}{2\pi}\right)^3\int e^{i \mathbf{k}\cdot\mathbf{y}} \mathbf{f}^*\left(\mathbf{r}+\frac{\mathbf{y}}{2},t\right)\cdot\mathbf{g}\left(\mathbf{r}-\frac{\mathbf{y}}{2},t\right)\mathrm{d}\mathbf{y}$, it is possible to derive a set of transport equations for $W_i$ \cite{bib:santos}: 
\begin{equation}
\partial_t W_0+\hat{\cal{L}}(W_2+W_3)=0 
\label{eq:dtW0}
\end{equation}
\begin{equation}
\partial_t W_1-\hat{\cal{G}}(W_2+W_3)-\frac{2}{\sqrt{\gamma_0}}\;W_2=0 
\label{eq:dtW1}
\end{equation}
\begin{equation}
\partial_t W_2-\hat{\cal{L}}W_0+\hat{\cal{G}}W_1+\frac{2}{\sqrt{\gamma_0}}\;W_1=0
\label{eq:dtW2}
\end{equation}
\begin{equation}
\partial_t W_3+\hat{\cal{L}}W_0-\hat{\cal{G}}W_1=0  
\label{eq:dtW3}
\end{equation}
where the operators $\hat{\cal{L}}$ and $\hat{\cal{G}}$ obey:
\begin{equation}
\hat{\cal{L}}= \sqrt{\gamma_0}\;\mathbf{k}\cdot\overrightarrow{\nabla}_\mathbf{r}-
\sqrt{\gamma_0}\; \left(\frac{n}{\gamma}\right)
\sin{\left(\frac{1}{2}\overleftarrow{\nabla}_\mathbf{r}\cdot\overrightarrow{\nabla}_\mathbf{k}\right)}
\label{eq:ops1}
\end{equation}
\begin{equation}
\hat{\cal{G}}=\sqrt{\gamma_0} \; \left(\mathbf{k}^2-\frac{\overrightarrow{\nabla}_{\mathbf{r}}^2}{4}\right)+\sqrt{\gamma_0}\; \left(\frac{n}{\gamma}\right)\cos{\left(\frac{1}{2}\overleftarrow{\nabla}_\mathbf{r}\cdot\overrightarrow{\nabla}_\mathbf{k}\right)}
\label{eq:ops2}
\end{equation}
 with the arrows denoting the direction of the operator, and $\sin(...)$ and $\cos(...)$ represent the equivalent series expansion of the operators; we observe that the left arrow operator $\overleftarrow{\nabla}_\mathbf{r}$ acts on $\sqrt{\gamma_0}\; \left(\frac{n}{\gamma}\right)$, while the right arrow operators ($\overrightarrow{\nabla}_\mathbf{k}$, $\overrightarrow{\nabla}_\mathbf{r}$, ${\overrightarrow{\nabla}_\mathbf{r}}^2$) act on $W_i$.  Equations (\ref{eq:dtW0}-\ref{eq:ops2}) are formally equivalent to the full wave equation for $\mathbf{a}$ in a plasma.

In order to close the system of Eqns.~(\ref{eqn:dens},~\ref{eq:dtW0}--\ref{eq:dtW3}), and to determine the corresponding dispersion relation, it is necessary to linearize Eqns.~(\ref{eq:dtW0}--\ref{eq:dtW3}), noting that up to first order $W_0=\sqrt{\gamma_0}\rho_0(\mathbf{k})\, \omega(\mathbf{k}) + \tilde{W}_0 (\mathbf{k}, \mathbf{r}, t)$, $W_1=\tilde{W}_1 (\mathbf{k}, \mathbf{r}, t)$, $W_2=-\rho_0(\mathbf{k}) \frac{\gamma_0 \mathbf{k}^2}{2}+\tilde{W}_2 (\mathbf{k}, \mathbf{r}, t)$, $W_3=\rho_0(\mathbf{k}) \left(1+\frac{\gamma_0 \mathbf{k}^2}{2} \right) +\tilde{W}_3 (\mathbf{k}, \mathbf{r}, t)$, 
where $\rho_0(\mathbf{k})=W_{\mathbf{a}_p\cdot\mathbf{a}_p}$ is the zero-order photon distribution function. In analogy with the standard techniques in plasma physics, $\rho_0(\mathbf{k})$ can be thought of as the equilibrium distribution function of the photons. Furthermore, we observe that $\tilde{W_2}+\tilde{W}_3=2W_{\mathrm{Re}\left[\mathbf{a}_p\cdot\mathbf{\tilde{a}}\right]}$. 
Linearization of Eqns.~(\ref{eq:dtW0}--\ref{eq:dtW3}), followed by time and space  Fourier transforms ($\partial_t \rightarrow \omega_L$, $\nabla_\mathbf{r} \rightarrow - i \mathbf{k}_L$), leads to:
\begin{equation}
{\cal F} \left[W_{\mathrm{Re}\left[\mathbf{a}_p\cdot\mathbf{\tilde{a}}\right]}\right]=\frac{1}{2} 
\, {\cal F}\left[\tilde{\left(\frac{n}{\gamma}\right)} \right] 
\left(\frac{\rho_0\left(\mathbf{k}+\frac{\mathbf{k}_L}{2}\right)}{D_s^{-}}+\frac{\rho_0\left(\mathbf{k}-\frac{\mathbf{k}_L}{2}\right)}{D_s^{+}}\right)
\label{eq:fwigner}
\end{equation}
where $D_s^{\pm}(\mathbf{k})=\omega_L^2\mp2\left[\mathbf{k}\cdot\mathbf{k}_L-\omega_L \, \omega\left(\mathbf{k}\mp\frac{\mathbf{k}_L}{2}\right)\right]$, and with ${\cal F}[g]_{\omega_L, \mathbf{k}_L}$ denoting the Fourier transform of $g(\mathbf{r},t)$.
In order to obtain the plasma response $\tilde{\left(\frac{n}{\gamma}\right)}$, the same technique is followed, leading to
\begin{equation} 
{\cal F} \left[ \tilde{\left(\frac{n}{\gamma}\right)}\right]=\frac{1}{\gamma_0^3}\left(\frac{\mathbf{k}_L^2}{\omega_L^2-\frac{1}{\gamma_0}}-1\right){\cal F} \left[\mathrm{Re}\left[\mathbf{a}_p\cdot\tilde{\mathbf{a}}\right] \right]
\label{eq:plasre}
\end{equation}
Integrating Eq.~(\ref{eq:fwigner}) in $\mathbf{k}$, 
and using $\int W_{\mathbf{f} \cdot \mathbf{g}} \, \mathrm{d} \mathbf{k} 
= \mathbf{f}^* \cdot \mathbf{g}$, Eqns.~(\ref{eq:fwigner},\ref{eq:plasre}) can then be combined to give the exact dispersion relation for electron plasma waves in the presence of broadband radiation:
\begin{equation}
1=\frac{1}{2\gamma_0^3}\left(\frac{\mathbf{k}_L^2}{\omega_L^2-\frac{1}{\gamma_0}}-1\right)
\int \rho_0\left(\mathbf{k} \right) \left(\frac{1}{D^+}+\frac{1}{D^-}\right) \mathrm{d}\mathbf{k}
\label{eq:gendispr}
\end{equation}
with $D^\pm(\mathbf{k}) = \left(\omega(\mathbf{k}) \pm \omega_L\right)^2 -\left(\mathbf{k} \pm \mathbf{k}_L\right)^2 -\frac{1}{\gamma_0}$. Equation~(\ref{eq:gendispr}) is the central result of this paper, and it generalizes the seminal result of Decker {\it et al} \cite{bib:decker} for pump fields with arbitrary statistics. It can be interpreted as the statistical average of 
$\frac{1}{D^+(\mathbf{k})}+\frac{1}{D^-(\mathbf{k})}$ over the distribution of photons.
For a pump plane wave, with wavenumber $\mathbf{k}_0$, $\rho_0(\mathbf{k})=a_0^2 \delta(\mathbf{k}-\mathbf{k}_0)$, and Eq.~(\ref{eq:gendispr}) leads to the same dispersion relation as derived in  Ref.~\cite{bib:decker}. Recently, a dispersion relation with two pump waves was also obtained \cite{bib:shukla06}, which also can be derived from Eq.~(\ref{eq:gendispr}) for two photon beams $\rho_0(\mathbf{k})=a_{0\,1}^2 \delta(\mathbf{k}-\mathbf{k}_{0\,1})+a_{0\,2}^2 \delta(\mathbf{k}-\mathbf{k}_{0\,2})$. 

To illustrate some of the most important consequences of white light in SRS, we consider the one dimensional scenario, for a water-bag zero-order photon distribution function $\rho_{0 \, \mathrm{WB}}(k)=a_0^2/(\sigma_1+\sigma_2)(\Theta(k-k_0+\sigma_1)-\Theta(k-k_0-\sigma_2))$, where $\Theta(k)$ is the Heaviside function. With this choice for $\rho_0$, several analytical results can be derived, highlighting the influence of white light in parametric instabilities. For $\rho_{0 \, \mathrm{WB}}(k)$, the random phase $\psi(x)$ is such that the autocorrelation function satisfies $\langle \exp{\left(-i\psi\left(x+\frac{y}{2}\right)+i \psi\left(x-\frac{y}{2}\right)\right)}\rangle=\exp \left(-i y \overline{k} \right) \sin{(y\overline{\sigma})}/(y \overline{\sigma})$, 
with $\overline{\sigma}=(\sigma_2+\sigma_1)/2$ and $\overline{k} =(k_0+(\sigma_2-\sigma_1)/2)$. The correlation length of this distribution is $\sim \pi/\sqrt{2} \overline{\sigma}$. 
The dispersion relation (\ref{eq:gendispr}) for this distribution function, valid for all values of $k_0$, $a_0$ and $\sigma_{1,2}$, is:
\begin{equation}
1=\frac{a_0^2}{8 \gamma_0^3k_L\overline{\sigma}}\left[\frac{k_L^2}{\omega_L^2-\frac{1}{\gamma_0}}-1\right] 
\left [ \frac{k_L^2}{k_L^2-\omega_L^2}\log{\left(\frac{D_1^-D_2^+}{D_1^+D_2^-}\right)}+
 \frac{2\omega_L k_L}{\sqrt{Q^0}}\left(
\mathrm{arctanh}\,{b^+} + \mathrm{arctanh}\,{b^-} \right) \right] ,
\label{eq:disp}
\end{equation}
where $\omega_{0i}=\omega\left(k_0+(-1)^i\sigma_i\right)$, 
$D^{\pm}_i=\omega_L^2 -k_L^2 \pm 2\left[(k_0+(-1)^i\sigma_i)k_L-\omega_{0i}\omega_L\right]$, 
$Q^{0}=\left(k_L^2-\omega_L^2\right)\left(k_L^2-\omega_L^2+\frac{4}{\gamma_0}\right)$,  
$Q^{\pm}=\left[D^{\pm}_1+(k_L-\omega_L) (\omega_L-2\omega_{01})\right] \left[D^{\pm}_2+(k_L-\omega_L) (\omega_L-2\omega_{02})\right] $,
 and $b^\pm = 2 k_L^2 (\omega_L+k_L) \sqrt{Q_0} \left( 2 \overline{\sigma}+ \omega_{01}-\omega_{02} \right)/\left( Q_0 k_L^2-Q^\pm (\omega_L+k_L)^2\right)$. 
 It is now possible to study the effect of a broadband photon distribution on stimulated Raman forward scattering (RFS), relativistic modulation instability (RMI) or stimulated Raman back scattering (RBS). 

Analytical results can be obtained in the case of an underdense medium $1/\gamma_0\ll k_0-\sigma_1$, which also guarantees that $k_0>\sigma_1$. The first condition states that the medium is underdense for all the photons in the distribution, while the second assures that $\rho_{0 \, \mathrm{WB}}(k)$ represents a broadband source of forward propagating photons. We further observe that under this approximation no order relation for $\sigma_2$ needs to be assumed. The underdense approximation is equivalent to neglecting the $\mathrm{arctanh}$ terms in (\ref{eq:disp}), since $\omega_{0i} \approx k_0+ (-1)^i\sigma_i$, and $b^\pm \approx 0$. We observe that by neglecting $1/\gamma_0$ when compared with $k_0$, the simplified dispersion relation is still valid for RBS, but we loose the ability to capture RMI. 
For RFS, we then Taylor expand the $\log$ term in Eq.~(\ref{eq:disp}), 
in the underdense approximation $k_0 -\sigma_1 \gg 1/\sqrt{\gamma_0}$. 
The resulting polynomial equation can the be evaluated near the wavenumber $k_L$ for maximum growth rate, such that $k_{L \, \mathrm{RFS} }^M \approx 1/\sqrt{\gamma_0}$, and $\omega_L \approx 1/\sqrt{\gamma_0}+\delta$, with $\delta \ll 1$, to yield the maximum growth rate for RFS, $\Gamma_{\mathrm{RFS}} = \mathrm{Im}[\delta]$:
\begin{equation}
\Gamma_{\mathrm{RFS}} = \frac{a_0}{2\sqrt{2} \gamma_0^2 \sqrt{\left(k_0-\sigma_1 \right) \left(k_0+\sigma_2 \right)}}.
\label{eq:grrfs}
\end{equation}
In the limit of $\sigma_{1,2} \rightarrow 0$, the standard monochromatic result, valid for all intensities, is obtained \cite{bib:decker}, with $\Gamma_{\mathrm{RFS}} \propto a_0^{-1}$ for $a_0 \gg 1$.  The effect of the bandwidth on four wave processes, such as RFS, predicted by Eq.~(\ref{eq:grrfs}) is qualitatively different from the effect of the radiation bandwidth in 3-wave processes, such as RBS. Eq.~(\ref{eq:grrfs}) shows that 
$\Gamma_{\mathrm{RFS}}$ increases(decreases) for increasing $\sigma_1$($\sigma_2$); this can be interpreted from the monochromatic result as due to the decrease(increase) of the average wavenumber of the distribution of photons. 
%
For RBS, the same technique can be followed, but now $D_2^+$ must be resonant (corresponding to the contribution of the downshifted photons of the highest wavenumber photons in the distribution function), with $\omega_L \approx 1/\sqrt{\gamma_0}+i \Gamma_\mathrm{RBS}$, with $\Gamma_\mathrm{RBS} \ll 1/\sqrt{\gamma_0}$ as in the usual treatment of RBS \cite{bib:kruer}, away from the strongly coupled regime, which means that the instability occurs in regions close to 
$k_{L \, \mathrm{RBS}}^M \approx 2 (k_0+\sigma_2)-1/\sqrt{\gamma_0}$. 
Furthermore, $D_1^+\simeq 0$, corresponding to the contribution of the downshifted photons of the lowest wavenumber photons in the distribution function, establishes the lower limit of the range of unstable wavenumbers given by 
$k_{L\, \mathrm{RBS}}^m \approx 2 (k_0-\sigma_1)-1/\sqrt{\gamma_0}$.
The maximum growth rate for RBS is:
\begin{equation}
\Gamma_{\mathrm{RBS}} = \frac{\pi a_0^2}{8 \gamma_0^{5/2}}
\frac{k_0+\sigma_2}{\sigma_1+\sigma_2} 
\frac{1}{1+\frac{a_0^2}{8 \gamma_0^{5/2}}
\frac{k_0+\sigma_2}{\left(\sigma_1+\sigma_2\right)^2}}
\label{eq:grrbs}
\end{equation}
where we have also assumed in the derivation of Eq.(\ref{eq:grrbs}) that $\Gamma_\mathrm{RBS} < \sigma_1+\sigma_2$. In the opposing limit, as $\sigma_{1,2} \rightarrow 0$ for the plane wave limit, 
we obtain $\Gamma_{\mathrm{RBS}\, \mathrm{pw}}=a_0 \sqrt{k_0}/\sqrt{2} \gamma_0^{5/4}$, thus recovering the standard result \cite{bib:kruer}.
For $a_0 \gg 1$, Eq.~(\ref{eq:grrbs}) shows the scaling $\Gamma_{\mathrm{RBS}} \propto a_0^{-1/4}$ as in the monochromatic case \cite{bib:decker}.
%
%
A comparison between Eq.~(\ref{eq:grrfs}) and Eq.~(\ref{eq:grrbs}) shows the stronger dependence of RBS on the bandwidth of the radiation, for $\sigma_2 \lesssim k_0$. In fact, for fixed $k_0$, $a_0$, and $\sigma_1$, RFS scales with $\propto 1/\sqrt{\sigma_2}$, while RBS goes as $\propto 1/\sigma_2$. 
We have found this behavior for other distribution functions (e.g. asymmetric lorentzian/gaussian distribution of photons), and, within the aproximations discussed here, it is possible to show that the scaling of RBS and RFS with $\sigma_2$ is independent of the exact shape of the distribution function.  
The wavenumber for maximum growth $k_L^M$ is independent of $\sigma_2$ for RFS, while for RBS it depends linearly on $\sigma_2$.  
Furthermore, Eq.~(\ref{eq:grrbs}) also shows that for large values of $\sigma_2$, the growth rates for RBS satures at $\Gamma_\mathrm{RBS}^\mathrm{sat} = \pi a_0^2/8 \gamma_0^{5/2}$.

To illustrate these features, we have solved Eq.~(\ref{eq:disp}) numerically, for the range of $k_L$ required to capture RFS, RBS, and RMI, and for different $\sigma_2$, keeping $k_0$, $a_0$, and $\sigma_1$ constant. 
In Figure~1, the dependence of the maximum growth rate with $\sigma_2$ are shown, demonstrating that RFS is less sensitive to $\sigma_2$ than RBS, as predicted by the theory. 
The theoretical curves, given by Eqns.~(\ref{eq:grrfs},\ref{eq:grrbs}), can not be distinguished from the numerical solution.
RMI depends more strongly on $\sigma_2$ than RFS, but still decreases more slowly than RBS, with $\Gamma_\mathrm{RMI}/\Gamma (\sigma_2=0) \simeq 1/\left(1+1.35(\sigma_2/k_0)^{1.1}\right)$, as obtained from a fit to the curve in Fig.~1. 
Figure~2 illustrates the main effects of the increase of $\sigma_2$ on RFS, showing that $k_{L \, \mathrm{RFS}}^M$ for maximum growth is almost independent of $\sigma_2$, and the decrease in the range of unstable $k_L$. The increase of the growth rate with $\sigma_1$, as predicted by Eq.~(\ref{eq:grrfs}), is also clear when we compare the growth rate for the monochromatic case, also shown in Fig.~2 (pw -- plane wave), with the other scenarios. 
This scenario is not observed for RBS (cf. Figure~3); RBS shows not only the decrease of maximum growth rate with $\sigma_2$ but also the increase of the range of unstable wavenumbers, and the shift of the most unstable wavenumber as $\sigma_2$ increases, in very close agreement with $k_{L\, \mathrm{RBS}}^M$. Furthermore, the lower bound of the range of unstable wavenumbers remains unchanged as $\sigma_2$ increases since this limit is determined by 
$k_{L\, \mathrm{RBS}}^m$, a function of $\sigma_1$. The magnitude of the growth rate is within the same order of magnitude for the full range of unstable wavenumbers which indicates, that depending of the noise source, the instability can easily grow in a wide range of wavenumbers. 
The trend of saturation of the growth rate with $\sigma_2$, as predicted for a waterbag distribution function, is also clear. 

We have studied electronic parametric instabilities in the presence of broadband fields; as expected, the maximum growth rates for these instabilities decrease with increasing $\sigma_2$. Nevertheless, a qualitative difference has been identified between RFS and RBS in the underdense limit, and for $\sigma_2 \lesssim k_0$, with RFS decreasing slower than RBS, thus less sensitive to broadband fields, illustrating a fundamental difference between these two processes. Numerical solutions of the generalized dispersion confirm the theoretical results and illustrate the increase of the range of unstable wavenumbers with $\sigma$.  
Generalization of this work to address the role of broadband radiation in self-focusing, stimulated Brillouin scattering is straightforward, thus allowing comparison with recent experimental results \cite{bib:montgomery}, and numerical particle-in-cell simulations. These results will be presented in future publications. 

\begin{acknowledgements}
The authors would like to acknowledge Prof. Warren Mori for valuable discussions, and Funda\c{c}\~ao para a Ci\^encia e a Tecnologia (Portugal) for financial support.
\end{acknowledgements}




\begin{figure}[h]
\begin{center}
\includegraphics[width=5.0 cm]{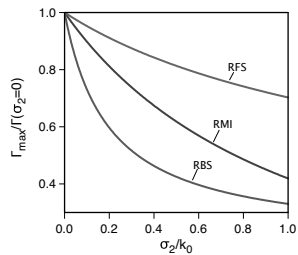}
\caption{Maximum growth rate as a function of the photon distribution width ($\sigma_2$), normalized to the growth rate for $\sigma_2=0$. The photon distribution is defined by $k_0=80 \, \omega_{p0}/c$, $a_0=0.1$ and $\sigma_1=0.2 \, k_0$.}
\label{fig:1}
\end{center}
\end{figure}

\begin{figure}[h]
\begin{center}
\includegraphics[width=5.0 cm]{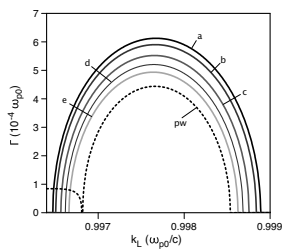}
\caption{Growth rate of the instability in the range of $k_L$ leading to RFS, for a photon distribution with $k_L$ $k_0=80 \, \omega_{p0}/c$, $a_0=0.1$ and $\sigma_1=0.2 \, k_0$, 
a -- $\sigma_2 = 0.29 \, k_0$, b -- $\sigma_2 = 0.39 \, k_0$, c -- $\sigma_2 = 0.59 \, k_0$, d -- $\sigma_2 = 0.78 \, k_0$, e -- $\sigma_2 = 0.98 \, k_0$. The plane wave (pw) pump scenario is obtained for the same parameters $k_0=80 \, \omega_{p0}/c$, $a_0=0.1$ -- RMI is also observed in this case.}
\label{fig:2}
\end{center}
\end{figure}

\begin{figure}[h]
\begin{center}
\includegraphics[width=5.0 cm]{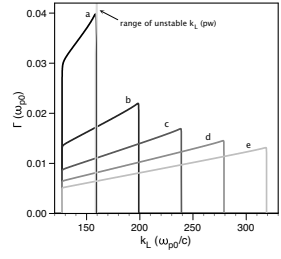}
\caption{Growth rate of the instability in the range of $k_L$ leading to RBS for a photon distribution with  $k_0=80 \, \omega_{p0}/c$, $a_0=0.1$ and $\sigma_1=0.2 k_0$, 
a -- $\sigma_2 = 0$, b -- $\sigma_2 = 0.25 \, k_0$, c -- $\sigma_2 = 0.5 \, k_0$, d -- $\sigma_2 = 0.75 \, k_0$, e -- $\sigma_2 =  k_0$. The range of unstable wavenumbers for the monochromatic scenario (pw) is also shown for reference, with a $\Gamma_\mathrm{RBS \, \mathrm{pw}}= 0.44 \, \omega_{p0}$. }
\label{fig:3}
\end{center}
\end{figure}


\begin{thebibliography}{99}
\bibitem{bib:mitchel} M. Mitchell \emph{et al}, Phys. Rev. Lett. \textbf{77}, 490 (1996); M. Mitchell and M. Segev, Nature (London) \textbf{387}, 880 (1997) 
\bibitem{bib:mori} W. B. Mori, IEEE Journal of Quantum Electronics \textbf{33}, 1942 (1997)
\bibitem{bib:buljan} D. N. Christodoulides \emph{et al}, Phys. Rev. Lett. \textbf{78}, 646 (1997); H. Buljan \emph{et al}, Phys. Rev. E \textbf{66}, 035601(R) (2002); H. Buljan \emph{et al}, Optics Lett. \textbf{28}, 1239 (2003); H. Buljan \emph{et al}, Phys. Rev. E \textbf{68}, 036607 (2003); B. Hall \emph{et al.}, Phys. Rev. E. \textbf{65}, 035602(R) (2002); D. Anderson \emph{et al.}, Phys. Rev. E. \textbf{69}, 025601 (2004); D. Anderson \emph{et al.}, Phys. Rev. E. \textbf{70}, 026603 (2004)
\bibitem{bib:kruer} W. L. Kruer, \emph{The Physics of Laser Plasma Interactions, 1st ed, 78-79} (Addison-Wesley Publishing Company Inc., 1988), and references therein
\bibitem{bib:thompson} C. Thompson \emph{et al.}, Astrophys. J. \textbf{422}, 304-335 (1994)
\bibitem{bib:mora} T. M. Antonsen and P. Mora, Phys. Fluids B \textbf{5}, 1440 (1993); P. Mora and T. M. Antonsen, Phys. Plasmas \textbf{4}, 217 (1997)
\bibitem{bib:zasvaleo} G. M. Zaslavskii and V. E. Zakharov, Sov. Phys. - Tech. Phys. \textbf{12}, 7 (1967); 
E. Valeo and C. Oberman, Phys. Rev. Lett. \textbf{30}, 1035 (1973)
\bibitem{bib:tsytovich} V. N. Tsytovich, \emph{Nonlinear Effects in Plasma} (Plenum, New York, 1970)  
\bibitem{bib:fisch} G. Shvets \emph{et al.}, Phys. Rev. Lett. \textbf{81}, 4879 (1998)
\bibitem{bib:liboff} R. L. Liboff, \emph{Kinetic Theory}, 2nd Ed., New York, 1998, p. 345
\bibitem{bib:tappert} I. M. Besieris and F. D. Tappert, J. Math. Phys. \textbf{14}, 704 (1973)
\bibitem{bib:pk} N. L. Tsintsadze and J. T. Mendon\c{c}a,
Phys. Plasmas \textbf{5}, 3609 (1998); J. T. Mendon\c{c}a and N. L. Tsintsadze, Phys. Rev. E \textbf{62}, 4276 (2000); L. O. Silva \emph{et al.}, IEEE Trans. Plasma Sci. \textbf{28}, 1202 (2000); R. Bingham \emph{et al.}, J. Plasma Phys. \textbf{71}, 899 (2005) 
\bibitem{bib:santos} J. P. Santos, L. O. Silva, J. Math. Phys. \textbf{46} , 102901 (2005)
\bibitem{bib:mckinstrie} C. Mckinstrie and R. Bingham, Phys. Fluids B \textbf{4}, 2626 (1992) 
\bibitem{bib:decker} C. D. Decker \emph{et al.}, Phys. Plasmas \textbf{3}, 2047 (1996)
\bibitem{bib:shukla06} P. K. Shukla \emph{et al.}, Phys. Plasmas \textbf{13}, 053104 (2006)
\bibitem{bib:scb} S. V. Bulanov \emph{et al.}, Phys. Fluids B \textbf{4}, 1935 (1992); P. Sprangle and E. Esarey, Phys. Rev. Lett. \textbf{67}, 2021 (1991); E. Esarey and P. Sprangle, Phys. Rev. A \textbf{45}, 5872 (1992); A. S. Sakharov and V. I. Kirsanov, Phys. Rev. E \textbf{49}, 3274 (1994) 
\bibitem{bib:montgomery} D. Montgomery, private communication (2006)
\end{thebibliography}
\end{document}